\documentclass[sn-mathphys-num,iicol]{sn-jnl}

\usepackage[T1]{fontenc}
\usepackage{lmodern}
\usepackage[utf8]{inputenc}
\usepackage[english]{babel}
\usepackage{graphicx}
\usepackage{hyperref}
\usepackage[dvipsnames]{xcolor}
\usepackage{color}
\usepackage{amsmath}
\usepackage{soul}
\usepackage{stix}
\hypersetup{colorlinks=true,citecolor={Mahogany},linkcolor={Mahogany},urlcolor={RoyalBlue}}
\usepackage{float}

\newcommand{\ket}[1]{\left| #1\right\rangle}
\newcommand{\bra}[1]{\left\langle #1\right|}

\newcommand\Tr{\mathrm{Tr}}

\usepackage{soul}
\usepackage[normalem]{ulem}

\unnumbered
\begin{document}

\title{Quantum and classical processing with photonic quantum
  machine learning}

\author*[1,2]{ J. C. López Carreño}
\email{juclopezca@cft.edu.pl}
\equalcont{}

\affil[1]{Center for Quantum-Enabled Computing, Center for
  Theoretical Physics of the Polish Academy of Sciences, Aleja
  Lotnik\'ow 32/46, 02-668 Warsaw, Poland}

\affil[2]{Institute of Experimental Physics, Faculty of Physics,
  University of Warsaw, ul. Pasteura 5, 02-093 Warsaw, Poland}

\author*[1]{S.~Świerczewski}
\equalcont{}
\email{swierczewski@cft.edu.pl}

\author[2,3]{A.~Opala}
\affil[3]{Institute of Physics of the Polish Academy of Sciences, Aleja Lotnik\'ow 32/46, 02-668 Warsaw, Poland}

\author[4]{A.~Salavrakos}
\affil[4]{Quandela, 7 Rue L\'eonard de Vinci, 91300 Massy, France}

\author[2]{B.~Pi\k{e}tka}

\author[1,3]{M.~Matuszewski}

\date{\today}

\abstract{Artificial intelligence and machine learning have been
  widely adopted both in the industry and in everyday life, but at
  the cost of high compute demands. Recent studies show that
  implementing machine learning in physical systems in the deep
  quantum regime could not only lead to faster information
  processing, but also to perform tasks that are out of reach for
  classical systems. Here, we report a quantum reservoir processing
  device capable of performing both quantum and classical machine
  learning tasks. The implementation is realized with a programmable
  silicon chip excited with single photons, a highly scalable and
  adaptable photonics technology. We successfully implement a
  variety of quantum tasks, including quantum state tomography and
  measurement of entanglement via negativity. Moreover, we implement
  a method of mitigation of experimental imperfections which results
  in a significant improvement in accuracy in comparison to the same
  system operating in the classical regime. Our results demonstrate
  a method to overcome a crucial bottleneck of quantum technologies
  by providing a practical way of probing quantum states. }

\maketitle

Machine learning implemented in deep neural networks is the most
versatile and impactful recent technology, with artificial
intelligence entering virtually all branches of economy and everyday
life. Quantum computing, on the other hand, has not yet developed
into a mature technology, but its potential to speed up practical
computations is believed to be extremely far-reaching. To date,
reported quantum advantage
demonstrations~\cite{Arute2019,Zhong2020,wu2021strong,madsen2022quantum},
remain limited to specific problems with no direct practical
relevance. Universal quantum computers promise broader applications,
but fault-tolerant computations require scale and gate fidelities
that are currently
inaccessible~\cite{Daley_PracticalQuantumAdvantage}.  Quantum
simulators provide proven quantum advantage
~\cite{Daley_PracticalQuantumAdvantage,Dziarmaga_BeyondClassical}
but remain application-specific and face model-simulator disparities
that are difficult to resolve.

Physical quantum machine learning (QML) is a less-known, but
promising alternative to quantum advantage, in principle applicable
for arbitrary machine learning tasks. QML leverages the complexity
of the Hilbert space and quantum effects to accelerate or improve
solutions of classical machine learning~\cite{Biamonte2017}. In
physical QML, quantum transformations are implemented natively in
hardware, rather than as a digital computing algorithm. This allows
for the performance of quantum tasks that are by definition
inaccessible for classical systems, such as tomography of a quantum
state or quantum state
generation~\cite{Ghosh_QuantumReservoir,ghosh2021quantum,ghosh2020reconstructing,Ghosh_StatePreparation,ko2025estimation}. This
is a particularly interesting angle for QML, as classical ML models
have become strikingly powerful, and it is often unclear where QML
models can outperform them for classical data. Intrinsically quantum
tasks, on the other hand, provide QML models with a natural
advantage. In general, there is a strong belief in the community
that QML models will be best suited for quantum
data~\cite{Huang2021}.

A prime example of physical QML is quantum reservoir computing
(QRC)~\cite{Fujii,Ghosh_QuantumReservoir}, with variants including
quantum reservoir processing (QRP) and quantum extreme machine
learning~\cite{Ghosh_QuantumReservoir,Pan_EnhancedImageRecognition}. QRC
relies on a physical quantum network called the reservoir, which
performs a nonlinear transformation of input data into a
high-dimensional feature space. This quantum reservoir is
non-trainable, but supplemented by a trainable linear readout layer,
which is often implemented in software. Recent studies show that QRC
achieves high performance on a variety of classical and quantum
tasks~\cite{Zambrini_ScalablePhotonicPlatform,nerenberg2025,sakurai2025,krisnanda2025experimental,ghosh2021quantum,swierczewski2026quantum,Krisnanda_Tomography,ghosh2021realising}. The
simplicity of implementation, minimal physical requirements and
system-agnostic learning rule make it currently one of the most
promising approaches for QML~\cite{Mujal_Opportunities}. Indeed,
training QML models is still considered a challenge: the
optimisation landscapes can suffer from barren plateaus and local
minima~\cite{McClean2018,Anschuetz2022} and the backpropagation
algorithm which made classical neural networks so successful cannot
be applied straightforwardly to quantum circuits. The community has
turned to alternatives which include classical-shadow-based
protocols~\cite{Huang2020}, train-on-classical, deploy-on-quantum
techniques~\cite{recioarmengol2026}, or QRC.

While superconducting circuits, neutral atoms, and trapped ions have
achieved strong positions in the quantum computing race, photonic
quantum computing offers distinct advantages: operation at optical
frequencies, robust room-temperature coherence, and compatibility
with existing fiber-optic communication
infrastructure~\cite{Pan_BosonSamplingwith20InputPhotons,Pan_Scalable_photonicquantumtechnologies,Steinbrecher_Quantumopticalneuralnetworks,Angelakis_Fockstateenhancedexpressivity}. Despite
the rapid development of QRC theory in recent years, experimental
realizations in the optical domain are scarce. Several recent works
demonstrated the possibility of improving model accuracy in the
quantum
regime~\cite{Rambach_PhotonicQuantumAccelerated,Parigi_Experimentalmemorycontrol,Gigan_HarnessingPhotonIndistinguishability,Paternostro_ExperimentalPropertyReconstruction,Paternostro_Quantumextremelearningmachinesforphotonicentanglement,McMahon_Largescalequantumreservoircomputing,negoro2018machinelearningcontrollablequantum,selimovic2025experimental}. However,
no demonstration of an optical system performing a quantum task has
been reported to date.

In this work, we report the experimental demonstration of a photonic
quantum reservoir system successfully processing quantum inputs and
realizing quantum tasks. We implement the protocol on an integrated
photonic quantum processing unit (QPU)~\cite{maring2024}. This
circuit, fabricated using a scalable silicon photonics technology,
is excited with a source of single photons. Crucially, it is capable
of performing photon number resolving (PNR) detection, while nearly
all previous experiments were limited to binary detection (0 or 1
photon). We implement a variety of tasks, including quantum
tomography of an entangled, mixed quantum state, using just a single
measurement basis. This is in stark contrast to conventional
tomography techniques, which in general require an exponential
number of measurements in different
bases~\cite{LvovskyCVTomography,elben2023randomized}. We extract key
quantum features of the input state, including its purity, von
Neumann entropy and negativity.

Apart from demonstrating quantum tasks, processing classical data
remains highly relevant for applications. Schemes have been
considered~\cite{nerenberg2025,
  sakurai2025,Rambach_PhotonicQuantumAccelerated} where classical
data is encoded in linear optical architectures used as
reservoirs. Here, we implement a technique that addresses
discrepancies between simulation and experiment.  This method
outperforms, in terms of accuracy, an ideal system operating in the
classical regime. Our results open the way to practical quantum
advantage in machine learning, achievable with scalable photonic
devices based on integrated photonic chips with single-photon
sources.

\section*{Results}

\subsection*{Photonic Quantum Processing Unit}
\begin{figure*}[hbt!]
    \centering
    \includegraphics[width=\textwidth]{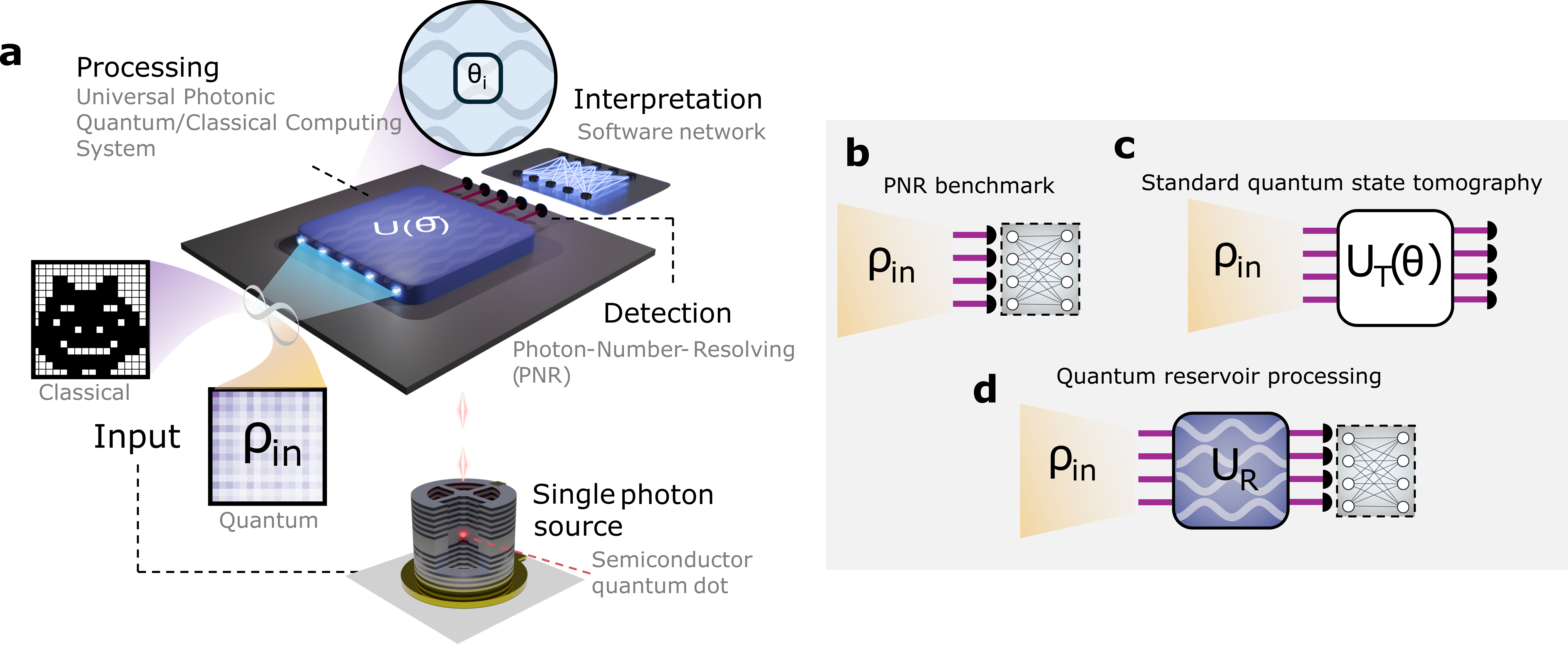}
    \caption{\textbf{Photonic quantum processing unit for quantum
        and classical machine learning tasks.}  \textbf{a}, Quantum
      reservoir processing with a programmable interferometric
      circuit. Classical or quantum input is encoded into a QPU and
      processed using single-photon states transformed by an
      arbitrary unitary $U(\bar{\theta})$, implemented with
      programmable parameters $\bar{\theta}$, which include phase
      shifts and coupling coefficients. Photon-number-resolving
      (PNR) detectors measure the output state photon-count
      probability distribution, which is further processed with a
      software neural network.  \textbf{b--d}, Approaches to quantum
      state tomography of the input state $\rho_{\rm in}$, including
      (\textbf{b}) direct PNR detection on the input state,
      (\textbf{c}) standard quantum state tomography with
      measurements in exponential number of different bases and
      (\textbf{d}) quantum reservoir processing using a single,
      fixed unitary
      transformation. 
    }
    \label{fig:fig_1}
\end{figure*}
The concept of the experimental realization is illustrated in Fig.~\ref{fig:fig_1}. 
The optical chip (QPU) consists of a mesh of integrated optical
waveguides arranged as an array of Mach-Zehnder Interferometers
(MZIs), which in turn are made of photonic mode couplers and thermal
phase shifters. The latter can be controlled with applied voltage,
thus allowing the QPU to implement arbitrary unitary operations.
This setup is implemented in Quandela's QPU chip ``Belenos'', which
supports up to 24 photonic modes, with up to 12 single photons
generated by a quantum dot source, simultaneously injected into a
subset of modes.
This initial state is a non-classical multimode Fock state, and the
coupling between modes in the QPU can give rise to mode
entanglement.  The output from the chip is analysed with
photon-number-resolving (PNR) detectors and an electronic
correlator, translating detector clicks into probabilities of
different multi-photon coincidences. These probabilities serve as
features constituting the input for the subsequent trainable
software neural network, which provides the final result of the
prediction.

In the case of classical tasks, the input data has typically a form
of a set of real-valued numbers. These are encoded directly in a
selected subset of QPU parameters, see Fig.~\ref{fig:fig_1}(a). On
the other hand, in the case of quantum tasks, the input has a form
of a quantum state. In this case, part of the QPU is devoted to
generating a quantum state, and the other part is used to perform
the task, cf.~Fig.~\ref{fig:tomography}. As an example,
Figs.~\ref{fig:fig_1}(b)-(d) illustrate three possible approaches to
the quantum tomography task. In the ``naive'' (PNR benchmark) case,
Fig.~\ref{fig:fig_1}(b), we use photon number resolving detectors
directly on the input state, followed by a software neural
network. This approach does not allow detecting any nonclassical
inter-mode correlations. In the standard quantum tomography,
Fig.~\ref{fig:fig_1}(c), a unitary transformation of the input state
before PNR is used to perform measurements in different bases. This,
however, requires an exponentially large number of different bases
to extract all elements of the density
matrix~\cite{LvovskyCVTomography}. Our QRP scheme,
Fig.~\ref{fig:fig_1}(d), combines the simplicity of single-basis
measurements with the ability to capture multimode quantum
correlations. It only requires a single, fixed reservoir
transformation and a trainable software neural network to
reconstruct the complete quantum
state~\cite{ghosh2020reconstructing}.

\subsection*{Quantum reservoir processing}

\begin{figure*}
    \centering
    \includegraphics[width=0.9\textwidth]{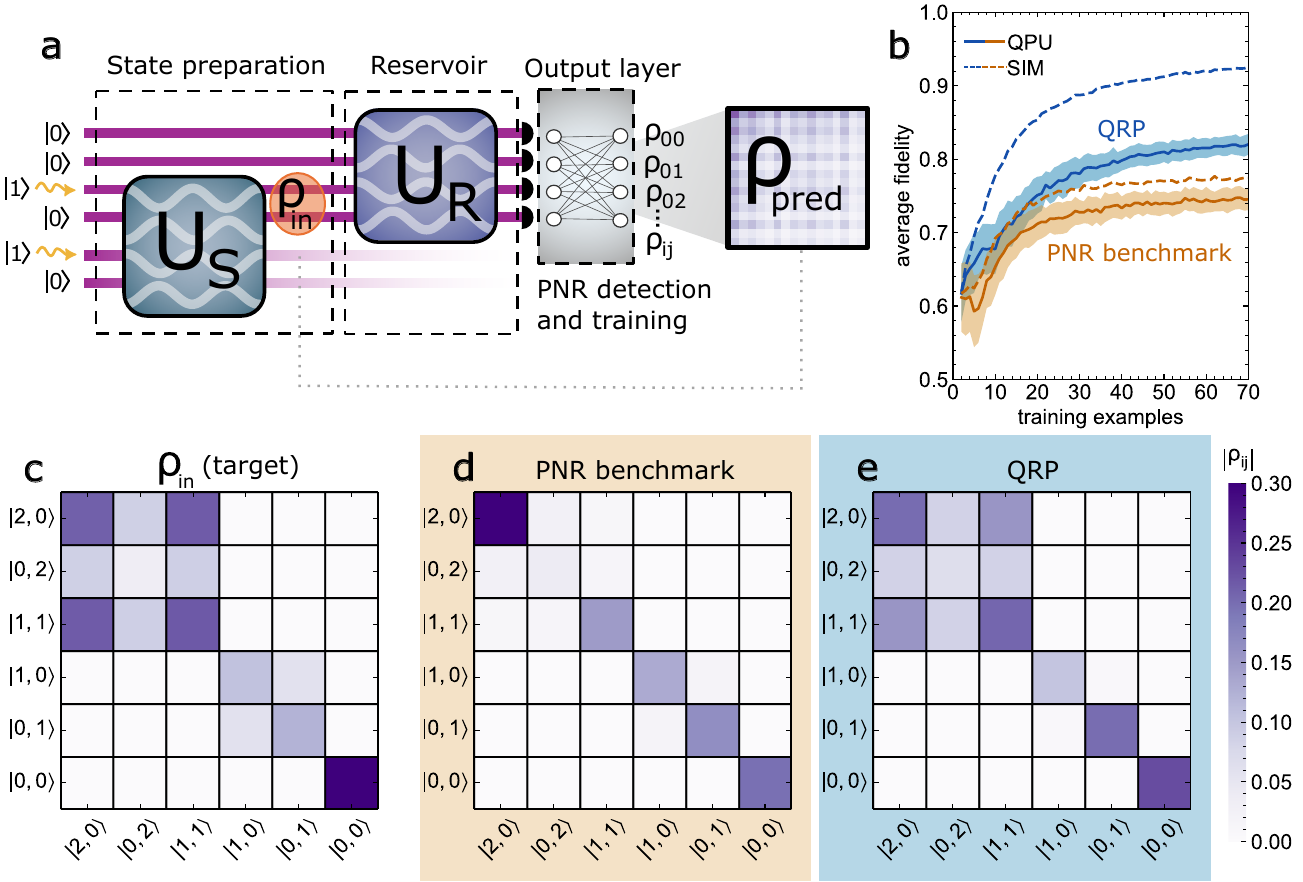}
    \caption{\textbf{Quantum state tomography of a two-photon mixed
        state.} \textbf{a} Schematic of the interferometric
      circuit. A pair of single photons in modes 3 and 5 is
      transformed by the unitary $\mathrm{U_{S}}$ acting on modes
      3–6. The resulting mixed input state $\rho_{\rm in}$ is
      generated in modes 3 and 4 by discarding ancillary modes 5 and
      6. This state is subsequently injected into the reservoir
      $\mathrm{U_{R}}$. The reconstructed density matrix
      $\rho_{\rm pred}$ is provided by a software neural network
      that analyzes the PNR coincidences at the reservoir output.
      \textbf{b} Average fidelity between $\rho_{\rm pred}$ and
      $\rho_{\rm in}$ in the testing dataset as a function of the
      number of training examples, obtained from the hardware QPU
      (solid lines) and a digital simulator (dashed lines). Blue
      lines correspond to QRP and orange lines to the PNR benchmark
      ($\mathrm{U_R} = \mathbb{I}$). Error bars show the standard
      deviation of result obtained over 50 different partitions into
      training and testing datasets.  \textbf{c–e} Density matrix
      representations showing (\textbf{c}) the target state,
      (\textbf{d}) the reconstruction performed with a PNR benchmark
      and (\textbf{e}) the QRP reconstruction. Note that QRP-based
      tomography recovers successfully the off-diagonal elements.}
    \label{fig:tomography}
\end{figure*}

\begin{figure*}[hbt!]
    \centering
    \includegraphics[width=0.8\linewidth]{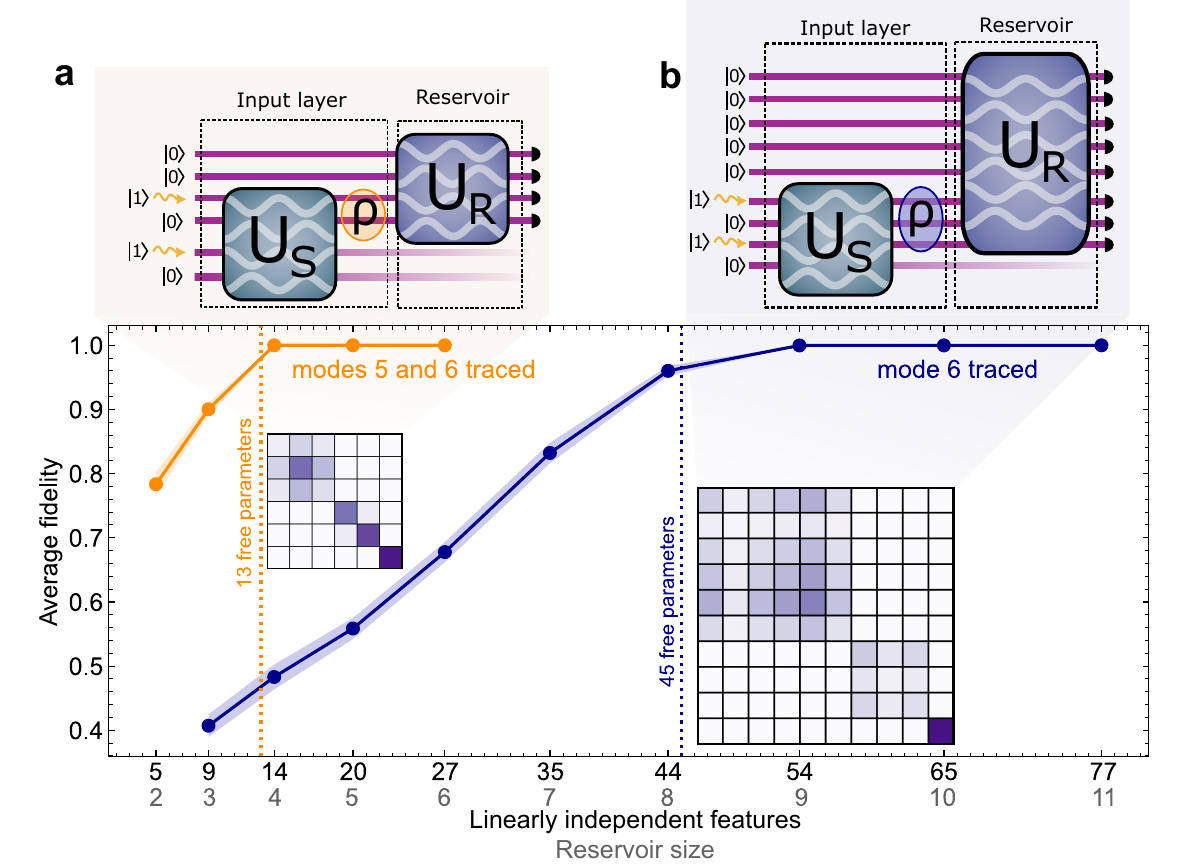}
    \caption{\textbf{Scalability of QRP-based quantum tomography.}
      Top: The interferometric circuit for (\textbf{a}) two- and
      (\textbf{b}) three-mode tomography. Bottom: Average tomography
      fidelity as a function of reservoir size and the number of all
      linearly independent photon-count configurations
      (coincidences) in the two- (orange line) and three-mode case
      (blue line). The shading corresponds to error bars obtained
      with 50 different choices of training and testing
      datasets. The insets depict the two- and three-mode density
      matrices together with the number of linearly independent
      real-valued parameters, which are indicated by orange (13) and
      blue (45) vertical dashed lines for the two- and three-mode
      states,
      respectively.}
    \label{fig:res_size}
\end{figure*}

As a proof-of-principle, we implemented the quantum tomography task
with the goal to reconstruct the density matrix of a mixed input
state $\rho$ in the two-mode basis, see~Fig.~\ref{fig:tomography}.
We programmed the initial state (Fig.~\ref{fig:tomography}a)
injected into the QPU interferometer as the two-photon state with
photons in modes 3 and 5,
$\ket{\psi}_{\mathrm{in}} = \ket{0,0,1,0,1,0,\ldots}$.  Crucially,
we postselect the output, keeping only the realisations where
exactly two photons are measured, effectively filtering out any
quantum trajectories containing loss events.

The initial state evolves according to the predefined state
preparation unitary, $\mathrm{U_{S}}$, acting on the subspace of
modes 3-6.  After this transformation, mixedness is introduced by
discarding ancillary modes 5 and 6, neglecting their measurement
outcomes.  This yields a reduced density matrix of the input state
$\rho_{\rm in}$ in modes 3 and 4 (see SI1 for the derivation) with a
block-diagonal structure
\begin{equation}
\rho_{\rm in} = \rho_{0}^{1\times1} \oplus \rho_{1}^{2\times2} \oplus \rho_{2}^{3\times3} ,\label{eq:mixed_state}
\end{equation}
where $\rho_{i}$ represents the subspace with $i$ photons left in modes 3 and 4 and the upper indices correspond to dimensions of the submatrices. We aim to reconstruct matrix elements of $\rho_{\rm in}$.

The mixed input state enters a reservoir defined by a fixed, random
$4 \times 4$ unitary, $\mathrm{U_{R}}$, acting on modes 1–4. This
transformation maps multimode quantum correlations into measurable
photon counting correlations via quantum interference. PNR detectors
collect coincidence probabilities for all possible measurement
outcomes, forming a 15-element feature vector.  At the output layer,
we performed the reconstruction using a software linear layer with
$\mathrm{L_{2}}$ regularization, followed by numerical matrix
renormalisation (see Methods and SI Sec.~1 for details).

The above QRP procedure was benchmarked against a pure PNR scheme
where $\mathrm{U_{R}}$ was replaced with the identity operation, see
Fig.~\ref{fig:fig_1}(b). In total, we experimentally implemented 100
quantum states, prepared by choosing random unitary
$\mathrm{U_{source}}$ matrices. The training labels were determined
by analytically calculating the corresponding density matrices
$\rho_{\rm in}$. From this dataset, 70 states were chosen at random
to construct the training dataset, with the remaining states used
for testing. As shown in Fig.~\ref{fig:tomography}\textbf{b}, the
reservoir consistently provided higher accuracy of predictions, both
in the QPU hardware implementation and in digital simulations
performed using the \textit{Perceval} Python
package~\cite{heurtel2023perceval}. The experimental QRP reached the
average fidelity on the testing dataset of
$F_{\rm QRP}=0.820 \pm 0.013$, outperforming the
$F_{\rm PNR}=0.747 \pm 0.015$ achieved by the PNR benchmark. The
reason for the improvement becomes clear when examining the density
matrix elements, with an example from the testing dataset depicted
in Fig.~\ref{fig:tomography}\textbf{d-f}. Without the reservoir, the
model captures diagonal elements (direct coincidence counts), but
fails to detect off-diagonal coherences due to the lack of
interference. The reported error of the average fidelity is
calculated by choosing different random training/testing dataset
splits. Since some states are easier to analyse than others, the
variance within a particular testing dataset can reach up to 10\% of
the reported average fidelity. In addition, based on the tomography
results we find Von Neumann entropy, negativity and purity of the
considered quantum states. These results are discussed in Sec.~1C of
SI.

\begin{figure*}
    \centering
    \includegraphics[width=\linewidth]{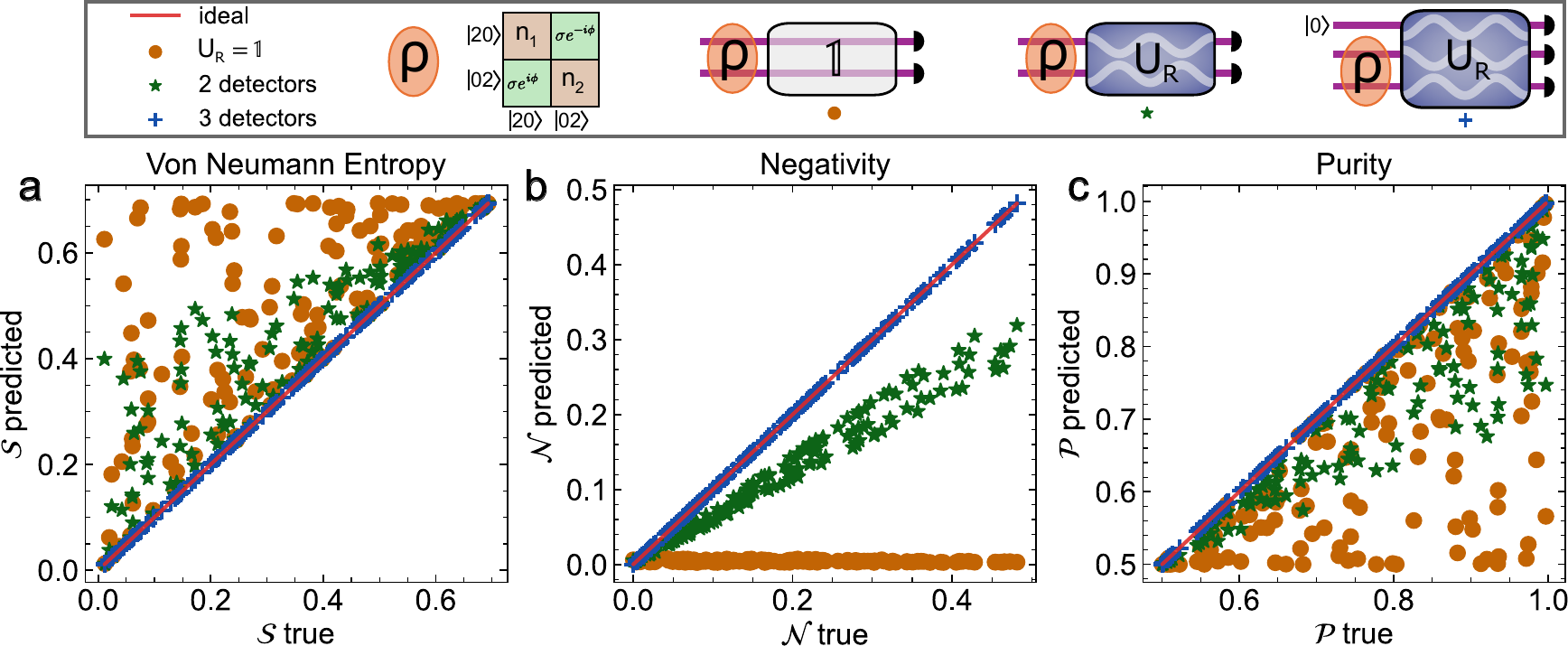}
    \caption{\textbf{Comparison between predicted and true values of
        quantum state characteristics.} \textbf{a} Von Neumann
      entropy, \textbf{b} negativity, and \textbf{c} purity of the
      input quantum state given by
      Eq.~(\ref{eq:ThuMar19142842CET2026}). In each panel, the three
      sets of points, correspond to the PNR Benchmark (orange
      circles), and QRP with two (green stars) and with three
      detectors (blue crosses). Solid red lines indicate the ideal
      prediction.}
    \label{fig:WedMar25153223CET2026}
\end{figure*}

To demonstrate the versatility and scalability of our approach to
QRP, we expand our model to perform quantum state tomography on a
three mode state, generated by tracing out one rather than two
modes. Using known results~\cite{ghosh2020reconstructing}, we
theoretically determine the minimal reservoir size, defined as the
necessary number of modes in the reservoir, i.e.~the dimension of
the unitary $\mathrm{U_{R}}$, required for successful tomography, in
both scenarios. The number of independent parameters can be
determined by examining the structure of the input density matrix,
as depicted in the insets in Fig.~\ref{fig:res_size}. In the two
mode case, considering all diagonal (real) and off-diagonal
(complex) parameters together with Hermiticity and unit trace
constraints results in 13 independent real-valued parameters,
whereas the three mode state has 45 independent parameters (see also
SI). This requires a correspondingly larger feature space to fully
determine the quantum state, as for a linear model to train
successfully, the number of linearly independent features,
representing distinct measurement outcomes, must be equal or exceed
the number of predicted parameters. We plot the average quantum
state tomography fidelity for both 2-mode and 3-mode cases against
the dimension of the feature space and the corresponding reservoir
size in Figs.~\ref{fig:res_size}\textbf{a}
and~\ref{fig:res_size}\textbf{b}.  A four- and nine-mode reservoir
are the minimum requirement for the two- and three-mode state,
respectively.  A deeper discussion about the scalability and
expressivity of our circuit is provided in the SI. There, we show
that the number of detectors needed to perform tomography grows
quadratically with the number of modes the state is defined on, with
a scaling factor depending only on the number of input photons. We
also considered classical tasks, finding that the rank of our
circuit transformation grows quadratically with the number of
features, scaled by a photon-number-dependent factor (Sec.~5 of the
SI).

While the experimental implementation considered here is limited to
Fock initial states and linear optical transformations, it is
important to investigate whether more complex quantum states can be
processed.  As an illustration, we consider the following two-photon
input state
\begin{multline}
    \label{eq:ThuMar19142842CET2026}
    \rho_\mathrm{in} = n_1 \ket{2,0}\bra{2,0} + n_2 \ket{0,2}\bra{0,2} +{}\\
    \sigma e^{i \phi}
    \ket{2,0}\bra{0,2} + \sigma e^{-i \phi}
    \ket{0,2}\bra{2,0},
\end{multline}
which is a valid physical state as long as $n_1 + n_2 =1$ and
$|\sigma| \leq \sqrt{n_1 n_2}$.  It is straightforward to calculate
analytically the probabilities of measuring the two photons at
certain outputs of the interferometer, see Sec.~2 of SI. These
probabilities are passed to a linear regression layer implementing a
tomography task. Based on the reconstructed density matrix we are
able to predict the value of three key quantum
characteristics~\cite{Ghosh_QuantumReservoir,ko2025estimation}: the
purity, the von Neumann entropy and the negativity of the quantum
state. Figure~\ref{fig:WedMar25153223CET2026} shows the prediction
accuracy in the case of the PNR benchmark, and the QRP scheme with
two and three detectors. In this case, training is performed with
1500 quantum states, while the testing dataset consists of 150
states, and each point shows the average over 50 different
reservoirs (see Methods for details). The results confirm the
advantage of QRP, and show that three detectors are already enough
to recover the input quantum state with very high fidelity. Again,
this can be understood by noting that the density matrix in
Eq.~(\ref{eq:ThuMar19142842CET2026}) has three independent
real-valued parameters and comparing this number with the number of
independent measurement outcomes for 2-detector (2) and 3-detector
(5) QRP.

One can also notice that since the negativity of the quantum state
in Eq.~(\ref{eq:ThuMar19142842CET2026}) is simply given by
$|\sigma|$, the PNR benchmark predicts $|\sigma| \approx 0$ in all
cases.  These considerations also apply to the expressivity of the
circuit: we find that the rank of the feature matrix is given by the
number of free parameters of the input density matrix \emph{plus}
one, and that in the collisionless boson sampling limit, rank grows
exponentially with the number of input photons (see Sec.~IV of the
SI).
Finally, we note that an alternative approach to the problem is to train the classical neural network to predict the three quantum characteristics directly, without the determination of the density matrix. However, we find that in this case the obtained accuracies are much lower (see SI).

\subsection*{Mitigation of experimental errors}

\begin{figure}
    \centering
    \includegraphics[width=1\linewidth]{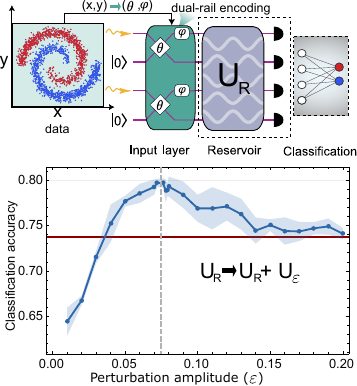}
    \caption{\textbf{Hardware-aware training for a classical task.}
      Random perturbations $U_{\rm \epsilon}$ in the reservoir
      matrix $U_{\rm R}$ with varying amplitudes $\epsilon$ are
      introduced in the in-silico training. This makes the system
      resistant to experimental imperfections.  Results are averaged
      over ten realisations of the reservoir matrix, and the shading
      indicates one standard deviation of the mean. The horizontal
      line shows the accuracy of an identical system in the
      classical regime, with coherent input states and average
      intensity measurement instead of PNR.}
    \label{fig:spiral}
\end{figure}

We implement a boson sampling architecture with classical inputs
encoded as interferometer parameters, see
Fig.~\ref{fig:spiral}. While similar systems have been considered
before~\cite{sakurai2025,Rambach_PhotonicQuantumAccelerated,nerenberg2025},
here we introduce a scheme that addresses the discrepancy between
simulation and experiment. As demonstrated below, with in-silico
training, an algorithm adapted to hardware imperfections is
necessary to obtain accuracies higher than those achievable with an
identical system operating in the classical regime.

In our training algorithm, we perturb the reservoir unitary
$U_{\rm R}$ by a small random unitary $U_{\rm \epsilon}$, different
for each sample. We modified each matrix element by a random value
drawn from a normal distribution with zero mean and $\epsilon^2$
variance. This can be considered as a form of regularization, in
which we train the network to become resistant to small deviations
of the experimental realization with respect to the idealized
case. For the results presented in this section, testing was
performed experimentally on the QPU.  As a proof of principle, we
consider a simple binary, nonlinear classification task, which is to
determine whether a specific point belongs to one of the two
intertwined spirals, see Fig.~\ref{fig:spiral}. Each point is
defined by a pair of coordinates $(x,y)$, so in principle a single
qubit could carry all the information. However, to exploit the full
capabilities of the setup, we introduce quantum interference by
encoding the data parameters in a pair of photons using the
dual-rail encoding scheme (see Sec.~4 of the Supplementary
Information). Then, we let them propagate through the reservoir
section of the interferometer, see~Fig.~\ref{fig:spiral}.

We used numerical simulations of the QPU to generate the training
dataset and to optimise the classical network (see Sec.~3 of the
Supplementary Information). To determine the experimental accuracy,
testing was performed on \emph{Ascella}, the QPU device preceding
\emph{Belenos}, operating with 12 photon
modes~\cite{maring2024}. This approach resulted in accuracy of
$\sim 79.7\%$, while a full \emph{in-silico} training and testing
strategy predicted over $98\%$ accuracy (cf. Fig.~SI4b). The reason
for such a discrepancy is the sensitivity of the training procedure
and the imperfect experimental implementation of the unitary matrix,
which leads to a discrepancy between programmed and physically
encoded unitaries. This effect can be mitigated to some extent by
including the perturbation during training, see
Figure~\ref{fig:spiral}, which shows how the accuracy of the
classification changes when random perturbations are included in the
elements of the unitary matrix. We find that fluctuation
amplitude~$\epsilon = 0.075$ provides the best accuracy given our
testing dataset, which is consistent with the measured transpilation
error in the \emph{Ascella} QPU. Only by using this training
technique were we able to surpass the accuracy of an ideal identical
classical network, where quantum input and PNR detection were
replaced with coherent state inputs and average intensity
measurements (horizontal line).

\section*{Discussion}

Quantum tomography and quantum feature extraction require in general
an exponential number of measurements in different bases, except for
specific input states or
observables~\cite{LvovskyCVTomography,elben2023randomized}.  We
demonstrate experimentally that QRP overcomes this crucial
bottleneck by performing a number-resolved measurement in a single
basis. This result has far-reaching consequences on applications,
including quantum computing, quantum sensing and quantum
communication technologies, where efficient probing of quantum
states is of great importance.  Most of these practical
implementations would require a somewhat modified setup. While here
the input state generation and tomography are performed inside a
single integrated device, input quantum states may be also generated
externally. Ideally, the training dataset would be labelled by
performing quantum tomography using an independent method.

A possible bottleneck in the scalability may result from the fact
that both the density matrix parameter space and the reservoir
feature space scale exponentially with the number of modes or
particles. This results in a mild scaling of reservoir size as shown
in Fig.~\ref{fig:res_size} and discussed in the SI, but can result
in an overwhelming workload for the classical network at the
output. However, in many practical applications, one only needs to
extract specific, useful quantum features rather than the full
density matrix~\cite{ko2025estimation,elben2023randomized}. This
allows for the reduction of feature space and better scaling for the
classical network (see SI).

\section*{Online Methods}

\subsection{Experimental details}

The \emph{Belenos} quantum processing unit follows an architecture
similar to that described in~\cite{maring2024}. It consists of a
single-photon source, a 12-mode active demultiplexer followed by
fibre delays, a 24-mode photonic integrated circuit made of silica
glass, and single-photon detectors. The single-photon source is a
quantum dot embedded inside a micropillar cavity, and is excited by
a laser to produce single photons on demand. At the time of the
experiment, single-photon purity (measured as the zero-delay
second-order correlation function, so the better the antibunching,
the purer the single photon emission) was estimated around
$(0.019 \pm 0.001)$ and photon indistinguishability around
$(0.923 \pm 0.007)$. The photons that are emitted are collected and
sent to the demultiplexer. The train of photons is converted by the
demultiplexer into up to 12 photons arriving simultaneously at
different modes of the chip.

The architecture of the chip is a universal Bell-Walmsley
interferometer~\cite{bell2021}, and its phase-shifters can be
controlled and tuned thermo-optically. Note that an $m$-mode
experiment with $m < 24$ can easily be implemented on a $24$-mode
chip by injecting photons only into chosen modes among the first
modes and blocking transmission into the last modes by controlling
the phases. In practice, the compilation of an experiment onto the
chip is done through an algorithm that is integrated into
\emph{Perceval}, and which employs machine learning
techniques~\cite{fyrillas2024}.

Finally, the photons reach the detectors, and detection events are
recorded via a time-to-digital converter. Importantly, the first 16
detectors corresponding to the first 16 modes of the chip have
polarization-resolved photon-number-resolving (PNR)
capabilities. The Belenos QPU is connected to a cloud platform,
which allows specific operations to be sent to the processor, and
which returns the results as detection statistics.

\subsection*{Quantum state tomography}
\label{subsec:QST_methods}

To reconstruct quantum states from QPU measurement data, we employ a
supervised machine learning approach to map vectors of photon
coincidence probabilities to corresponding density matrices. A
$d \times d$ complex density matrix $\rho$ is completely
characterized by $d^2 - 1$ independent real parameters, constrained
by Hermiticity ($\rho = \rho^\dagger$) and unit trace
($\text{Tr}(\rho)=1$). We construct a parameter vector by extracting
$d-1$ independent real diagonal elements alongside the real and
imaginary components of the strictly upper off-diagonal elements. To
satisfy the trace-one condition, the final diagonal element is
calculated as $\rho_{dd} = 1 - \sum_{i=1}^{d-1} \rho_{ii}$.

We train a multi-output ridge regression model implemented in
\textit{scikit-learn}, utilizing an $L_2$ regularization parameter
$\alpha = 10^{-3}$, to learn the linear mapping from the empirical
coincidence probability distributions to these target parameter
vectors. During inference, the model predicts a real vector that is
algebraically reshaped into an initial matrix $\tilde{\rho}$. To
enforce Hermiticity, we apply the transformation:
\begin{equation}
    \rho_{\text{H}} = \tilde{\rho} + \tilde{\rho}^\dagger - \text{diag}(\tilde{\rho})
\end{equation}
where $\text{diag}(\tilde{\rho})$ is a matrix containing only the
diagonal elements of $\tilde{\rho}$.

Because linear regression does not inherently respect the
semi-positivity constraint, $\rho_{\text{H}}$ may possess negative
eigenvalues. To obtain a physical density matrix prediction
$\rho_{\text{phys}}$, we perform an eigendecomposition
$\rho_{\text{H}} = V \Lambda V^\dagger$ and project the eigenvalues
onto the probability simplex. We set
$\lambda_i' = \max(\lambda_i, 0)$ and re-normalize the spectrum such
that $\sum \lambda_i' = 1$. The reconstructed state is then
recovered via:
\begin{equation}
    \rho_{\text{phys}} = V \text{diag}(\lambda') V^\dagger
\end{equation}

The accuracy of our reconstructions is evaluated using quantum fidelity
\begin{equation}
    \label{eq:WedApr1135643CEST2026}
    F(\rho_{\text{phys}}, \rho_{\text{target}}) = \left( \text{Tr} \sqrt{\sqrt{\rho_{\text{phys}}} \rho_{\text{target}} \sqrt{\rho_{\text{phys}}}} \right)^2\,,
\end{equation}
which gives us a measure, bounded by 0 and 1 of how well the state
is predicted, with increasing fidelity meaning better prediction.

For the second quantum task, we analysed some properties of a
two-mode, two-photon mixed state, as given in
Eq.~(\ref{eq:ThuMar19142842CET2026}). The learning and testing
datasets used in the analysis above were obtained using the
following algorithm: we draw three random numbers, $n$, $\gamma$ and
$\varphi$ from a uniform distribution in the interval $[0,1]$. Then,
we transformed these random numbers into parameters of the quantum
state as $n_1 = 1-n$, $n_2 = n$, $\sigma = \gamma \sqrt{n_1 n_2}$
and $\phi = 2\pi \varphi$. We used the transformation of a general
interferometer (cf. Section SI2 of the Supplemental Information for
the details) to compute the probability distribution of the photon
coincidences at the output of the interferometer. Finally, we
implemented a linear regression model using \emph{scikit-learn} to
determine the relation between the coincidence probability
distribution and the Bloch vector~$\vec{a}$ associated to the input
quantum state in Eq.~(\ref{eq:ThuMar19142842CET2026}); namely, the
three-dimensional vector~$\vec{a}$ with components
\begin{subequations}
    \begin{align}
        a_x &= 2 \sigma \cos \phi\,,\\
        a_y &= 2 \sigma \sin \phi\,,\\
        a_z &= n_1 - n_2\,,
    \end{align}
\end{subequations}
satisfying the relation~$\rho = (I + \Vec{a}\cdot \vec{\sigma})/2$,
where $I$ is the identity matrix and $\vec{\sigma}$ is the vector of
Pauli matrices. Using the recovered density matrix, we computed
three quantum features of interest, namely the purity
\begin{equation}
    P(\rho) = \Tr(\rho^2)\,,
\end{equation}
the von Neumann entropy
\begin{equation}
    S(\rho) = -\Tr(\rho \mathrm{ln} \rho) = - \sum_j \lambda_j
    \mathrm{ln} \lambda_j\,, 
\end{equation}
where~$\lambda_j$ are the eigenvalues of the density matrix~$\rho$,
and the negativity
\begin{equation}
  \mathcal{N}(\rho) = \frac{||\rho^{\Gamma_1}||_1-1}{2} = \sum_j
  \frac{|\eta_j|-\eta_j}{2}\,, 
\end{equation}
where $||\bullet ||_1$ is the trace norm, $\rho^{\Gamma_1}$ is the
partial transposition of the density matrix~$\rho$ with respect to
the subsystem 1, and $\eta_j$ are the eigenvalues of such a
partially transposed matrix.  For our quantum state in
Eq.~(\ref{eq:ThuMar19142842CET2026}) these read, respectively,
\begin{subequations}
\begin{align}
    P(\rho) & = n_1^2 + n_2^2 +2 \sigma^2\,,\\
    S(\rho) & =  \log 2 - \frac{1}{2} \log (1 - R^2) + \frac{R}{2} \log\left(\frac{1-R}{1+R} \right)\,,\\
    \mathcal{N}(\rho) &= |\sigma|
\end{align}
\end{subequations}
where~$R = \sqrt{(n_1-n_2)^2 + 4 \sigma^2}$; furthermore, it is easy
to show that these functions have the following bounds
\begin{subequations}
\begin{align}
    \frac{1}{2} \leq &\, P(\rho)  \leq 1\,,\\
    0 \leq & \, S(\rho) \leq  \log 2 \,,\\
    0 \leq &\, \mathcal{N}(\rho)  \leq \frac{1}{2}\,.
\end{align}
\end{subequations}
Namely, the purity is bounded from below by $1/d$, where $d$ is the
dimension of the Hilbert space in which the state is defined; the
von Neumann entropy runs from 0 for a pure state to $\log 2$ for a
maximally mixed state; and the negativity of general NOON states of
the form $\alpha \ket{N,0} + \beta \ket{0,N}$ is given by
$|\alpha \beta|\leq 1/2$, independently of $N$.
As shown in Fig.~\ref{fig:WedMar25153223CET2026}, the prediction
based on the probability distribution from three detectors is very
precise, and we do not need to make any post-processing of the
results neither to maximize the fidelity
[cf. Eq.~(\ref{eq:WedApr1135643CEST2026})] between the input and the
reconstructed quantum state, nor to recover perfectly the quantum
features of interest.

\section*{Acknowledgments}

This project received funding from the European Union’s Horizon
Europe research and innovation programme under grant agreement No
101130384 (QUONDENSATE). The Center for Quantum-Enabled Computing
project is carried out within the International Research Agendas
programme of the Foundation for Polish Science co-financed by the
European Union under the European Funds for Smart Economy 2021-2027
(FENG). A.O. acknowledges the project No. 2024/52/C/ST3/00324 funded
by the National Science Center, Poland.

\bibliography{sci}

\end{document}